\documentclass[twocolumn,pre,showpacs,showkeys]{revtex4}

\usepackage{amsmath}
\usepackage{amssymb,amsfonts}
\usepackage{bm}
\usepackage[mathcal]{euscript}
\usepackage{graphicx}
\usepackage{psfrag}

\newcommand{\comment}[1]{}

\newcommand{\mathnotation}[2]{\newcommand{#1}{\ensuremath{#2}}}

\renewcommand{\l}{\left}			
\renewcommand{\r}{\right}			
\mathnotation{\pd}{\partial}			
\mathnotation{\imi}{{\mathrm{i}}}		
\mathnotation{\ldef}{\mathrel{\raisebox{.069ex}{:}\!\!=}}
\mathnotation{\rdef}{\mathrel{=\!\!\raisebox{.069ex}{:}}}
\mathnotation{\unitI}{\mathbb{I}}               

\renewcommand{\time}{t}				
\mathnotation{\pdt}{\pd_\time}
\mathnotation{\xc}{x}				
\mathnotation{\xv}{{\bm{\xc}}}			
\mathnotation{\uc}{u}				
\mathnotation{\uv}{{\bm{\uc}}}		        

\mathnotation{\apolyc}{A}                       
\mathnotation{\apoly}{\mathbb{\apolyc}}         
\mathnotation{\apolyeq}{\apolyc_0}              
\mathnotation{\rpoly}{\rho}
\mathnotation{\rpolyc}{\rpoly}
\mathnotation{\rpolyv}{\bm{\rpolyc}}
\mathnotation{\rpolyeq}{\rpoly_0}               
\mathnotation{\rpolym}{\rpoly_{\mathrm{m}}}     
\mathnotation{\Fpolyc}{F}                       
\mathnotation{\Fpoly}{\bm{\Fpolyc}}             
\mathnotation{\visc}{\nu}                       
\mathnotation{\relt}{\tau}                      
\mathnotation{\De}{\mathrm{De}}                 
\mathnotation{\Z}{\mathcal{Z}}
\mathnotation{\PDF}{\mathcal{P}}
\mathnotation{\edirv}{\bm{e}}
\mathnotation{\frel}{f}
\mathnotation{\lyap}{\lambda}
\mathnotation{\lyapt}{\bar\lambda}
\mathnotation{\coil}{{\mathrm{c}}}              
\mathnotation{\stretched}{{\mathrm{s}}}         
\mathnotation{\taucr}{\Fpolyc_{\mathrm{c}}}     
\mathnotation{\rpolycr}{\rpoly_{\mathrm{c}}}    
\mathnotation{\Betaf}{B}                        
\mathnotation{\detr}{\mathcal{D}}               
\mathnotation{\Ldet}{\mathcal{L}}               

\begin{document}

\title{Finite Extension of Polymers in Turbulent Flow}
\author{Jean-Luc Thiffeault}
\email{jeanluc@mailaps.org}
\affiliation{Department of Applied Physics and Applied Mathematics, Columbia
University, New York, NY 10027}
\altaffiliation[Present address: ]{Department of Mathematics, Imperial College
  of Science, Technology, and Medecine, London SW7 2BZ.}

\begin{abstract}
The statistics of polymers advected by a turbulent flow are investigated.  To
limit the polymer lengths above to coil--stretch transition, a FENE-P type
relaxation law is used.  The turbulence is modeled by a random strain,
delta-correlated in time and with Gaussian statistics.  The distribution of
polymer lengths for both the coiled and stretched states are derived, from
which are obtained analytical expressions for the moments of the distribution.
The polymer stress on the fluid decreases linearly with the inverse of the
Deborah number.  The degradation (breaking) of the polymers is also discussed,
showing that even for large Deborah number some polymers can remain unbroken.
\end{abstract}


\maketitle

One of the principal motivations to study the properties of polymers diluted
in a turbulent flow is to understand the Toms effect, also known as drag
reduction~\cite{Lumley1969,Virk1975}, whereby the effect of a small
concentration of polymers is to dramatically reduce the pressure required to
maintain a turbulent flow at a given velocity in a pipe.  It is thought that
the extension of the polymers by the strains in the turbulent flow (the
coil--stretch transition) is crucial to that
effect~\cite{Lumley1969,Hinch1977}.  More recently, there has been great
progress in understanding the microscopic behavior of biological polymers such
as DNA~\cite{Perkins1994,Perkins1994b,Quake1997}, which give confidence that
there is a regime where the polymer tension dominates the relaxation
time~\cite{Hatfield1999}, as opposed to hydrodynamic effects.

The dynamics of polymers in incompressible turbulence was studied recently in
Refs.~\cite{Chertkov2000,Balkovsky2000,Balkovsky2001,Groisman2001}.
\Citet{Balkovsky2000,Balkovsky2001}, building on the approach of
\citet{Lumley1972}, modeled the turbulence as a homogeneous, isotropic,
delta-correlated random strain---the Kraichnan--Kazantsev model of
turbulence~\cite{Kraichnan1968,Kazantsev1968,Shraiman1994,Chertkov1995,%
Chertkov1998,Balkovsky1999} (see the review by \citet{Falkovich2001})---for
polymers whose relaxation is described by the Hookean dumbbell
model~\cite{Bird2}.  \Citet{Chertkov2000} extended the analysis to general
nonlinear relaxation models.  Experiments~\cite{Groisman2001} and numerical
simulations~\cite{Eckhardt2002} validate many of the findings of these papers.

In this letter we explore the coiled and stretched states of polymers assuming
the popular FENE-P model~\cite{Bird2,Ilg2002,DeAngelis2002} (Finite Extension
Nonlinear Elastic)
for the relaxation force of the polymer.  The advantage of the FENE model is
that its nonlinear nature prevents the polymers from being infinite in length
above the coil--stretch transition: it limits them to a finite, fixed length.
The results for the PDF of the polymer lengths in both the coiled and
stretched states are presented, as well as moments of the distribution, the
polymer stress on the fluid, and a discussion of polymer degradation by
breaking.  For definiteness and simplicity, Gaussian statistics for the strain
are assumed throughout, though the calculation herein could be redone for more
arbitrary statistics using the path integral formalism used by
\citet{Chertkov2000}.  The emphasis is thus on presenting closed form,
analytically simple expressions that give a reasonable description of the
physics.  Compressibility effects are neglected, as is the feedback of the
polymers on the flow, justified when the polymer stress is much less
than~$\visc\lyapt$~\cite{Balkovsky2001}, where~$\visc$ is the fluid
(Newtonian) viscosity and~$\lyapt$ is the mean rate of strain.

\section{Fokker--Planck Equation}
\label{sec:FP}

The shape of the polymers is characterized by the symmetric conformation
tensor,~$\apoly$, which describes the deformation of the polymers from a
coiled ball of radius~$\rpolyeq$ into a stretched ellipsoid in the presence of
flow.  We take the equation of motion for the conformation tensor to be that
of a general viscoelastic fluid with a relaxation force term (only the longest
restoring time is taken into account),
\begin{equation}
  \frac{d\apolyc^{ij}}{d\time}
  = \apolyc^{ik}\,\pd_k\uc^j + \pd_k\uc^i\apolyc^{kj}
  - \frac{2}{\relt} \l(\frel(\apolyc^{kk})\apolyc^{ij} -
  \rpolyeq^2\,\delta^{ij}\r),
  \label{eq:A}
\end{equation}
where~$\frel(\apolyc^{kk})$ depends only on the trace~$\apolyc^{kk}$
of~$\apoly$, and is equal to~$1$ when~$\apoly=\rpolyeq^2\,\unitI$,
and~\hbox{$d/dt = \pd/\pd\time + \uc^k\pd_k$} is the advective derivative.  We
take the velocity field to be incompressible,~$\pd_k\uc^k=0$.  (Unless
otherwise noted, repeated indices are summed.)  The conformation
tensor~$\apoly$ can be diagonalized, with eigenvalues~$\rpolyc_\alpha^2$ that
evolve according to
\begin{equation}
  \frac{d\rpolyc_\alpha}{d\time} = \hat\sigma_{\alpha\alpha}\,\rpolyc_\alpha
  - \frac{1}{\relt}\l(\frel(\rpolyc_\beta\rpolyc_\beta)\rpolyc_\alpha -
  \rpolyeq^2/\rpolyc_{\alpha}\r),
  \label{eq:rhoceq}
\end{equation}
with~$\hat\sigma_{\alpha\beta}(\time,\xv) \ldef
\edirv_\alpha^i\,\pd_k\uc^i\,\edirv^k_\beta$ the velocity gradient tensor
projected on the orthonormal eigenvectors,~$\edirv_\alpha$.  There is no sum
over~$\alpha$ on the right-hand side of~\eqref{eq:rhoceq}.

When the viscous scale of the flow is much larger than the polymer length (the
typical situation), the turbulence appears locally as a constant strain.  In
the context of our model, the polymers are stretched ellipsoids whose major
axis is aligned with the dominant stretching direction, associated with the
largest Lyapunov exponent, positive here for an incompressible flow.  This
alignment is observed in numerical experiments~\cite{Ilg2002}.  We thus
neglect the subdominant directions in~\eqref{eq:rhoceq} and write
\begin{equation}
  \frac{d\rpoly}{d\time} = \lyap\rpoly
  - \frac{1}{\relt}\l(\frel(\rpoly^2)\rpoly -
  \rpolyeq^2/\rpoly\r),
  \label{eq:rhoeq}
\end{equation}
where~$\rpoly$ is the largest~$\rpoly_\alpha$ and~$\lyap(\time)$ is the
Lyapunov exponent associated with the dominant direction of stretching.

We take~$\lyap(\time)$ to be a random variable satisfying
\begin{equation}
  \langle\lyap(\time)\lyap(\time')\rangle - \lyapt^2
  = \delta(\time - \time') \,\Delta;
  \qquad
  \langle\lyap(\time)\rangle = \lyapt,
\end{equation}
where the angle brackets denote a average over~$\lyap(\time)$.  In general we
have~\hbox{$\Delta \le \lyapt$}, since the standard deviation cannot be larger
than the mean for the positive Gaussian variable~$\lyap$.

To get a Fokker--Planck equation for the PDF of~$\rpoly$, we introduce the
characteristic function, $\Z(\time;\mu) =
\l\langle\exp\l(\imi\mu\rpoly\r)\r\rangle$.  We can then derive an equation of
motion for~$\Z$ from~\eqref{eq:rhoeq} and average.  Gaussian integration by
parts~\cite{Furutsu1963,Novikov1964,Donsker1964,Frisch} allows evaluation of
terms of the form~$\langle\lyap\Z\rangle$, finally yielding a closed equation
for~$\Z$.  Inverse Fourier transformation of~$\Z$ with respect to~$\mu$ then
gives the equation of motion for~$\PDF(\time,\rpoly)$, the PDF of~$\rpoly$,
\begin{equation}
  \pdt\PDF = \tfrac{1}{2}\Delta\,\pd_\rpoly\rpoly\,\pd_\rpoly\rpoly\,\PDF
  - \lyapt\,\pd_\rpoly\,\rpoly\,\PDF
  + \frac{1}{\relt}\,\pd_\rpoly\l(\frel(\rpoly^2)\rpoly -
  \rpolyeq^2/\rpoly\r)\PDF.
  \label{eq:FP}
\end{equation}
(The~$\pd_\rpoly$ act on all terms to their right.)  This is the same
Fokker--Planck equation as derived by Chertkov~\cite{Chertkov2000}, except
that instead of diffusivity we shall use the minimum polymer size~$\rpolyeq$
as an ultraviolet cutoff.  This also allows us to treat the coiled state in
Section~\ref{sec:coiled}.  \comment{And except for the sign error
in~\cite{Chertkov2000}, immaterial to the steady-state.}

\section{The Coiled State}
\label{sec:coiled}

We first deal with the case of relatively unstretched polymers, so that their
effective radius (the maximal extension of the ellipsoid describing the
deformation of the polymers) is much less than the maximal extension of the
polymer.  In that case we assume that the polymers are in a linear force
response regime and we can use the Hookean dumbbell
model,~\hbox{$\frel(\rpoly^2)=1$}.  Since the determinant of~$\apoly$ is
greater than~$(\rpolyeq^3)^2$ (see the appendix), we can assume that the
largest principal axis~$\rpoly$ (the only one we model) is larger
that~$\rpolyeq$, and assume the PDF of~$\rpoly$ vanishes
for~\hbox{$\rpoly<\rpolyeq$}.

The Fokker--Planck equation~\eqref{eq:FP} with~$\frel=1$ can be solved for the
steady-state distribution,
\begin{equation}
  \PDF_\coil(\rpoly) \sim
       \rpoly^{-1 - 2(\xi - \zeta)}\,\exp(-\xi\,\rpolyeq^2/\rpoly^2),
       \label{eq:PDFcoil}
\end{equation}
where~$\xi \ldef 1/\Delta\relt$, $\zeta \ldef \lyapt/\Delta$.  From the PDF we
can derive an exact expression for its moments,
\begin{equation}
  \frac{\langle\rpoly^n\rangle}{\rpolyeq^n}
  = \xi^{n/2}\,\frac{\gamma(\xi - \zeta - n/2,\xi)}{\gamma(\xi - \zeta,\xi)},
  \label{eq:rhoncoil}
\end{equation}
where~$\gamma(a,x)$ is the lower incomplete gamma function.
($\gamma(a,\infty)$ is equal to $\Gamma(a)$, the complete gamma function.)
For~$\xi>\zeta$, the denominator in~\eqref{eq:rhoncoil} is positive and
finite, and for~\hbox{$n < 2(\xi-\zeta)$} so is the numerator.

For~\hbox{$n = 2(\xi-\zeta)$} the numerator diverges, so that the
moment~$\langle\rpoly^n\rangle$ becomes infinite.  When~$\xi=\zeta$ the~$n=0$
moment diverges, which means that the PDF~$\PDF_\coil$ becomes unnormalizable.
Thus the physical picture is as follows: for large~\hbox{$\xi-\zeta$} only
moments associated with large~$n$ diverge, but as~\hbox{$\xi-\zeta$} is
lowered more moments diverge, indicating that on average more polymers have
anomalously long lengths, that is, they are uncoiled.  For~$\xi=\zeta$ all
positive ($n>0$) moments are infinite, which indicates that the polymers are
essentially all in a stretched (uncoiled) state, but also pointing to a
breakdown in the theory because of the infinite moments.

\begin{figure}
  \psfrag{Dei}{\raisebox{-1ex}{{$\De^{-1}$}}}
  \psfrag{rhon}{\raisebox{2ex}{\hspace{-1em}
      {$\langle\rpoly^n\rangle/\rpolyeq^n$}}}
  \centering\includegraphics[width=.9\columnwidth]{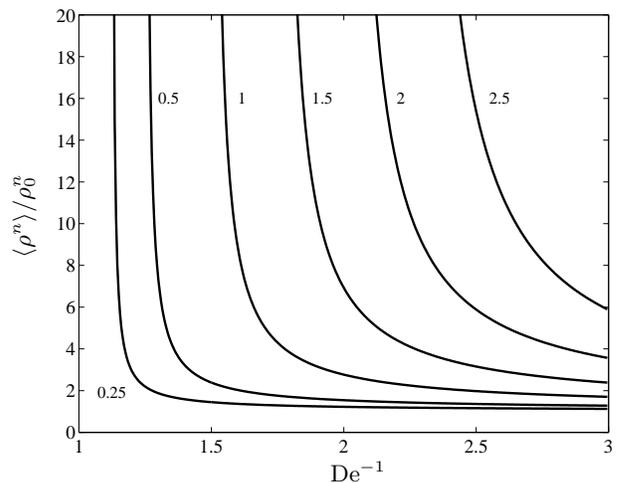}
  \caption{Moments~$\langle\rpoly^n\rangle/\rpolyeq^n$ of the PDF of polymer
  lengths in the coiled state for different values of~$n$,
  with~\hbox{$\zeta\protect\ldef\lyapt/\Delta=1$}.}
  \label{fig:rhoncoiled}
\end{figure}

The combination~$\xi-\zeta$ can be rewritten
\begin{equation}
  \xi - \zeta = \zeta(\De^{-1} - 1);\qquad \De \ldef \lyapt\,\relt,
\end{equation}
where~$\De$ is the Deborah number, the ratio of the polymer relaxation
timescale over the advection timescale.  The transition to the stretched state
thus occurs for~\hbox{$\De\ge1$}.  A few moments are plotted as a function of
the Deborah number in Fig.~\ref{fig:rhoncoiled}, showing their divergence
as~$\De$ approaches unity.

For~\hbox{$\zeta+1$} moderately large (fluctuations are small compared to the
mean),
the main contribution to the integral representation of the incomplete gamma
function occurs well below~$\xi$, so we can replace the upper integration
bound~$\xi$ by~$\infty$ in~\eqref{eq:rhoncoil} and obtain
\begin{equation}
  \frac{\langle\rpoly^n\rangle}{\rpolyeq^n}
  \simeq \xi^{n/2}\,\frac{\Gamma(\xi - \zeta - n/2)}{\Gamma(\xi - \zeta)}.
  \label{eq:rhoncoilsimp}
\end{equation}
The even moments then reduce to simple products of monomial factors according
to the factorial property of the gamma function.  A simplification occurs
for~$n=2$,
\begin{equation}
  \frac{\langle\rpoly^2\rangle}{\rpolyeq^2}
  \simeq \frac{\xi}{\xi - \zeta - 1}
  = [1 - \Delta(\lyapt + \relt)]^{-1}.
  \label{eq:rhoncoilsimp2}
\end{equation}
This is a particularly significant moment, since the polymer stress on the
fluid is proportional to~\hbox{$\apoly \sim \rpoly^2$}.

Note that the singular denominator in~\eqref{eq:rhoncoilsimp2}, which comes
from the gamma function numerator of~\eqref{eq:rhoncoilsimp}, implies that the
polymer length can be much larger than~$\rpolyeq$ even below the coil--stretch
transition.  Of course, when it becomes too large we must include nonlinear
contributions to the relaxation force, the topic of the next section.

\section{The Stretched State}
\label{sec:stretched}

As we saw in Section~\ref{sec:coiled}, the Hookean dumbbell model fails once
the Deborah number reaches unity, because it cannot prevent the polymers from
becoming infinitely long~\cite{Hinch1977}.  One standard way to prevent this
is to use the FENE-P model~\cite{Bird2,Ilg2002,DeAngelis2002}, for which
\begin{equation}
  \frel(\rpoly^2) = \frac{\rpolym^2 - \rpolyeq^2}{\rpolym^2 - \rpoly^2}\,.
  \label{eq:FENErel}
\end{equation}
The restoring force of the polymers is now such as to limit their length
to~$\rpolym$ (though they make break before reaching that length, as we see
below).  We assume that the elongation of the polymers is smaller than the
viscous length, so that their length is limited by their own elasticity and
not by the hydrodynamic mechanism of Tabor and de
Gennes~\cite{Tabor1986,DeGennes1986}.

Since in this section we are interested in the stretched state of long
polymers, we may neglect~\hbox{$\rpolyeq \ll \rpolym$} both
in~\eqref{eq:FP} and~\eqref{eq:FENErel}, which leads to the stationary
distribution
\begin{equation}
  \PDF_\stretched(\rpoly) \sim
       \rpoly^{-1 - 2(\xi - \zeta)}(1 - \rpoly^2/\rpolym^2)^\xi
       \label{eq:PDFstrech}
\end{equation}
Unlike the distribution obtained in~\cite{Chertkov2000,Balkovsky2000}, this
PDF exhibits a cutoff at~$\rpolym$, reflecting the fact the limited size of
the polymers.  We also impose a cutoff at~$\rpoly_0$ for small~$\rpoly$,
because~$\rpoly$ cannot be less than the equilibrium length~$\rpolyeq$ (see
the appendix).  In Ref.~\cite{Chertkov2000} diffusivity was used to regularize
the PDF at small~$\rpoly$, but both methods are equivalent since~$\rpolyeq$
drops out in our approximation.

The moments of the PDF~\eqref{eq:PDFstrech} are found to be
\begin{equation}
  \frac{\langle\rpoly^n\rangle}{\rpolym^n} =
  \frac{\Gamma(\zeta - \xi + n/2)\,\Gamma(\zeta + 1)}
       {\Gamma(\zeta - \xi)\,\Gamma(\zeta + 1 + n/2)}
       = \frac{{(\zeta - \xi)}_{n/2}}{{(\zeta + 1)}_{n/2}},
  \label{eq:rhonstrech}
\end{equation}
where~\hbox{$(x)_n \ldef \Gamma(x+n)/\Gamma(x)$} is the Pochhammer symbol.
This expression is valid for~\hbox{$\zeta-\xi>0$}, or equivalently
for~$\De>1$.  At~$\De=1$, Eq.~\eqref{eq:rhonstrech} predicts zero length for
the polymers, but this is just a symptom of the neglect of~$\rpolyeq$.
Rather, the coiled solution of Section~\ref{sec:coiled} must then be used.

\begin{figure}
  \psfrag{Dei}{\raisebox{-1ex}{{$\De^{-1}$}}}
  \psfrag{rhon}{\raisebox{2ex}{\hspace{-1em}
      {$\langle\rpoly^n\rangle/\rpolym^n$}}}
  \centering\includegraphics[width=.9\columnwidth]{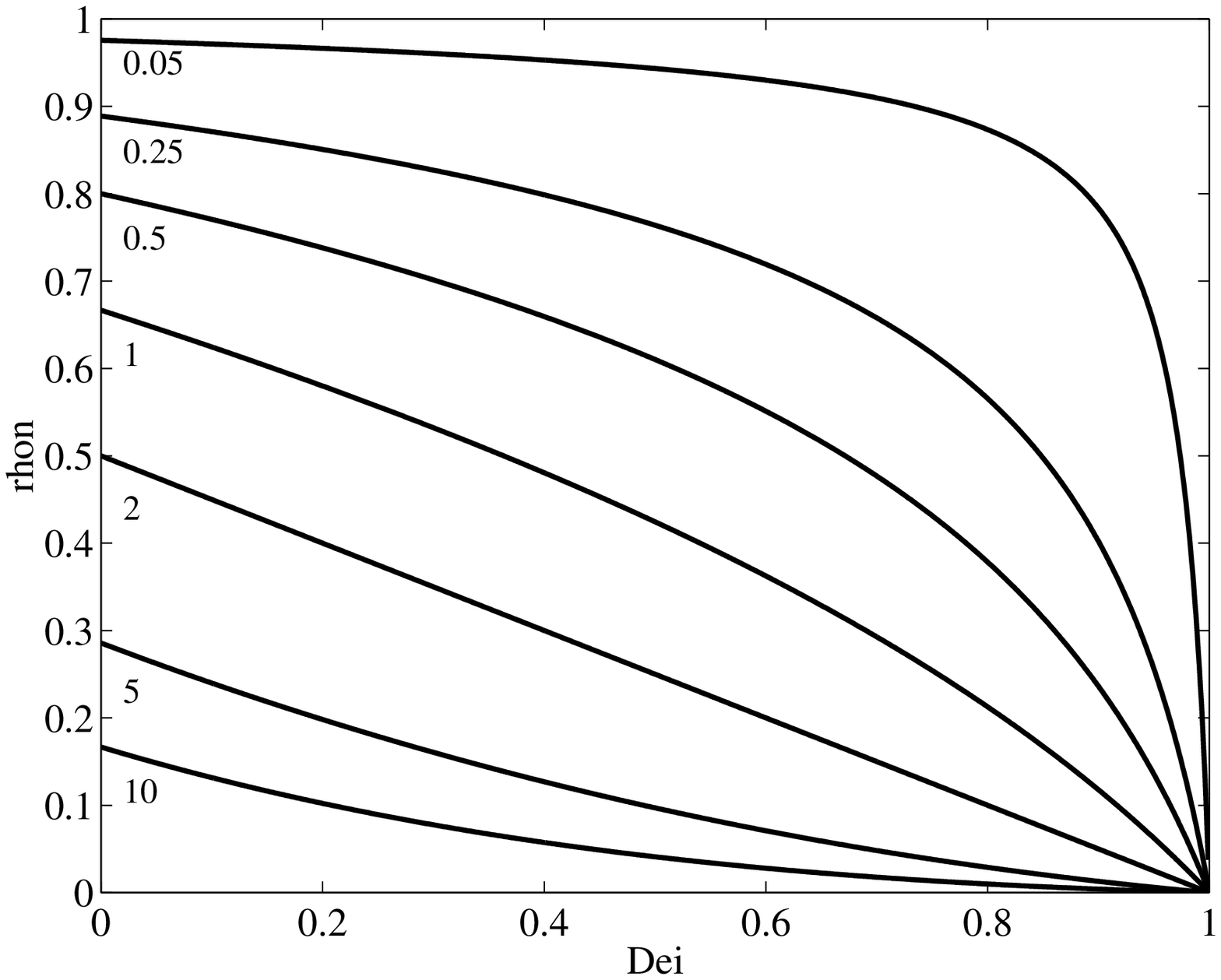}
  \caption{Positive moments~$\langle\rpoly^n\rangle/\rpolym^n$ of the PDF of
  polymer lengths in the stretched state,
  with~\hbox{$\zeta\protect\ldef\lyapt/\Delta=1$}.}
  \label{fig:rhonstretched}
\end{figure}

\begin{figure}
  \psfrag{Dei}{\raisebox{-1ex}{{$\De^{-1}$}}}
  \psfrag{rhon}{\raisebox{2ex}{\hspace{-1em}
      {$\langle\rpoly^n\rangle/\rpolym^n$}}}
  \centering\includegraphics[width=.9\columnwidth]{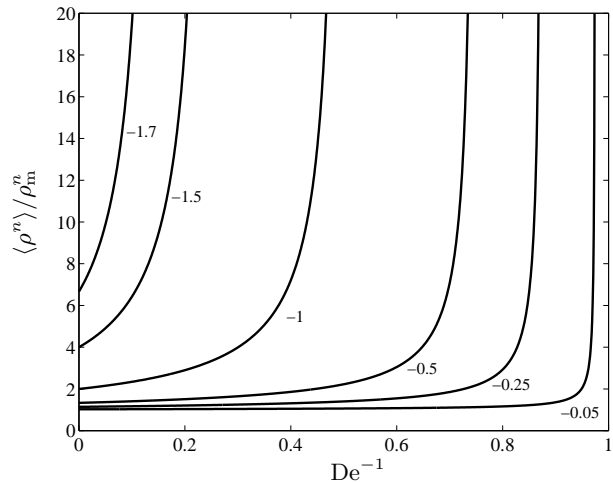}
  \caption{Negative moments~$\langle\rpoly^n\rangle/\rpolym^n$ of the PDF of
  polymer lengths in the stretched state,
  with~\hbox{$\zeta\protect\ldef\lyapt/\Delta=1$}.}
  \label{fig:rhonstretchedneg}
\end{figure}

For $n=2$, Eq.~\eqref{eq:rhonstrech} again takes a simplified form,
\begin{equation}
  \frac{\langle\rpoly^2\rangle}{\rpolym^2} =
  \frac{\zeta - \xi}{\zeta + 1}
  = \frac{\lyapt - \relt^{-1}}{\lyapt + \Delta}
  = \frac{1 - \De^{-1}}{1 + \zeta^{-1}},
  \label{eq:rhonstrechsimp2}
\end{equation}
so the average polymer stress (proportional to~$\langle\rpoly^2\rangle$) is
linear in the inverse Deborah number.  Some positive ($n>0$) and negative
($n<0$) moments are plotted as a function of the Deborah number in
Figs.~\ref{fig:rhonstretched}--\ref{fig:rhonstretchedneg}.
The negative moments diverge as~$\De$ approaches unity, a consequence of
the neglect of~$\rpolyeq$.

As mentioned in Ref.~\cite{Chertkov2000}, the PDF of length can also be used
to estimate the fraction of polymers that break. (This is a kind of
degradation, though not the same as say for DNA, where the two strands become
separate~\cite{Choi2002}.)  Assume a polymer breaks if the
tension~$\Fpolyc(\rpoly^2) = \frel(\rpoly^2)\rpoly/\relt$ exceeds a critical
value,~$\taucr$.  To that critical tension corresponds a critical
length,~$\rpolycr$, obtained by solving
\begin{equation}
  \taucr = \frac{1}{\relt}\,\frel(\rpolycr^2)\rpolycr = \frac{1}{\relt}\,
  \frac{\rpolym^2}{\rpolym^2 - \rpolycr^2}\,\rpolycr,
\end{equation}
a quadratic in~$\rpolycr$.
The fraction of polymers that survive is equal to the fraction shorter
than~$\rpolycr$, obtained by integrating the PDF~\eqref{eq:PDFstrech} from~$0$
to~$\rpolycr$,
\begin{equation}
  \text{Prob}\,(\rpoly < \rpolycr)
  = \frac{\Betaf(\rpolycr^2/\rpolym^2\,;\,\zeta-\xi\,,\,\xi+1)}
  {\Betaf(\zeta-\xi\,,\,\xi+1)},
  \label{eq:fracnbreak}
\end{equation}
where~\hbox{$\Betaf(a,b) \ldef \Gamma(a)\Gamma(b)/\Gamma(a+b)$} is the beta
function, and~$\Betaf(z;a,b)$ is the incomplete beta function.
Since~\hbox{$\Betaf(1;a,b) = \Betaf(a,b)$}, for~\hbox{$\rpolycr=\rpolym$} none
of the polymers break.
\begin{figure}
  \psfrag{Dei}{\raisebox{-1ex}{{$\De^{-1}$}}}
  \psfrag{Pnobreak}{\raisebox{2ex}{\hspace{-1em}
      {$\text{Prob}\,(\rpoly < \rpolycr)$}}}
  \centering\includegraphics[width=.9\columnwidth]{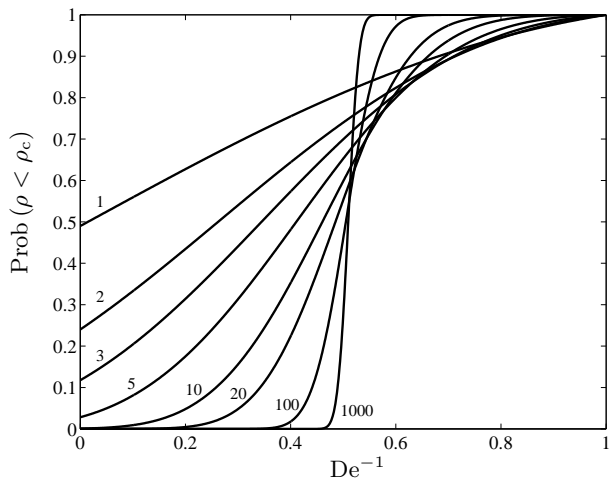}
  \caption{Fraction of polymers that survive as a function of the inverse
    Deborah number, for different values
    of~\hbox{$\zeta\protect\ldef\lyapt/\Delta$}
    with~\hbox{$\rpolycr/\rpolym=0.7$}.}
  \label{fig:Pnobreak}
\end{figure}
The fraction of surviving polymers~\eqref{eq:fracnbreak} is plotted in
Fig.~\ref{fig:Pnobreak} for different values of~$\zeta$.  For large~$\zeta$
(small fluctations), the survival probability becomes a step function with
``knee'' at~$\De^{-1} = 1 - (\rpolycr/\rpolym)^2$: either all polymers survive
or all are broken.

A consequence of~\eqref{eq:fracnbreak} is that a significant fraction of
polymers can survive even for large Deborah number; For~\hbox{$\xi\ll \zeta$}
(\hbox{$\De\gg 1$}) and~\hbox{$\xi\ll 1$}, \hbox{$\text{Prob}\,(\rpoly <
\rpolycr) \simeq (\rpolycr/\rpolym)^{2\,\zeta}$}, which is not small
if~$\zeta$ is not too large.  In other words, some polymers survive breakage
for large~$\lyapt$ if the fluctuations in~$\lyapt$ (given by~$\Delta$) are
also large.  Of course, this treatment is for one correlation time of the
turbulence.  In reality a given polymer is likely to be exposed to many
different random strains.  Because we have assumed the turbulence to be
delta-correlated, the probability of survival of the polymers decays roughly
as~$(\rpolycr/\rpolym)^{2\zeta N}$,
where~\hbox{$N=\time/\time_{\mathrm{corr}}$} with $\time$ the elapsed time
and~\hbox{$\time_{\mathrm{corr}}$} the correlation time. (More generally, for
smaller Deborah number, take the~$N$th power of~\eqref{eq:fracnbreak}.)

\appendix

\section*{Appendix: Lower Bound on Polymer Volume}

We show that the evolution equation~\eqref{eq:A} for the polymers implies a
lower limit on the volume of the polymer ellipsoid.  That volume is
proportional to~\hbox{$\det\apoly = \rpoly_1^2\rpoly_2^2\rpoly_2^2$}, so we
let~\hbox{$\detr\ldef\rpoly_1\rpoly_2\rpoly_3$}.  (We rescale~$\rpoly$ such
that~$\lVert\rpolyeq\rVert=1$.)  For small volumes (low stretching) the
Hookean approximation applies, so we let~$\frel=1$ in~\eqref{eq:rhoceq};
the quantity~$\log\detr$ obeys
\begin{equation}
  \relt\,\frac{d}{d\time}\log\detr =
  -3 + \l(\rpolyc_1^{-2} + \rpolyc_2^{-2} + \rpolyc_3^{-2}\r)
  \rdef \Ldet(\rpoly).
  \label{eq:dtdet}
\end{equation}
\begin{figure}
  \psfrag{det > 1}{{$\detr > 1$}}
  \psfrag{det < 1}{{$\detr < 1$}}
  \psfrag{det > 1,  L > 0}{{$\detr > 1,\  \Ldet > 0$}}
  \psfrag{L > 0}{{$\Ldet > 0$}}
  \psfrag{L < 0}{{$\Ldet < 0$}}
  \psfrag{rho1}{\raisebox{-2ex}{{$\rpoly_1$}}}
  \psfrag{rho3}{\raisebox{2ex}{{$\rpoly_3$}}}
  \centering\includegraphics[width=.7\columnwidth]{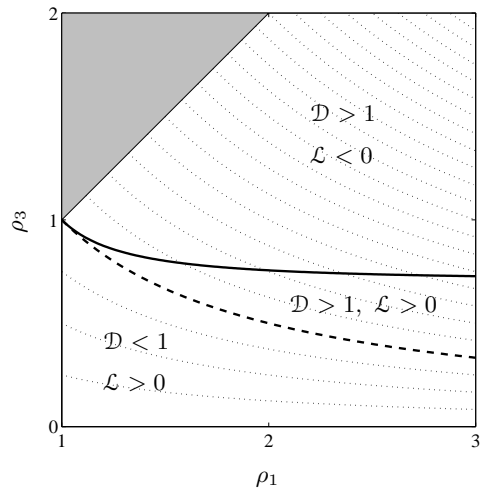}
  \caption{Stability diagram showing
  that~$\detr\protect\ldef\rpolyc_1\rpolyc_2\rpolyc_3$ must have value greater
  than or equal to unity.  Constant-$\detr$ curves are dotted, with
  the~$\detr=1$ curve dashed.}
  \label{fig:detstab}
\end{figure}
The sign changes of the derivative of~$\log\detr$ are the same as those
of~$\detr$ since the logarithm is monotonically increasing.
Figure~\ref{fig:detstab} shows the sign of~$\Ldet$ in
the~$\rpoly_1$--$\rpoly_3$ plane, assuming~$\rpoly_2=1$.  Without loss of
generality we also assume~\hbox{$\rpoly_1\ge\rpoly_3$}, so we need not
consider the shaded area.  The solid curve is~$\Ldet=0$, where the determinant
is constant.  The dotted curves show lines of constant~$\detr$, the dashed
line being~$\detr=1$.  The dotted and dashed curves meet at the spherical
solution.  Below the solid curve,~$\detr$ must increase, and above it must
decrease.  The line~$\detr=1$ lies entirely below~$\Ldet=0$ and is tangent to
it at~\hbox{$\rho_1=\rho_3=1$}.  Assuming that the polymers are initially in
the spherical (rest) state, we see that~$\detr$ must either remain at or
increase away from unity.  In fact the only disturbances that do not linearly
decay back to the spherical state are those along the solid line.
The picture for~\hbox{$\rpoly_2\ne 1$} is three-dimensional but is
qualitatively the same.  We conclude that~$\detr\ge1$, so that the volume of
the ellipsoid is no smaller than the equilibrium sphere.  Hence, the largest
polymer dimension,~$\rpoly_1$, is never smaller than~$\rpolyeq$, justifying
its use as a small-$\rpoly_1$ cutoff.

\begin{acknowledgments}
This work was supported by the National Science Foundation and the Department
of Energy under grant No.~DE-FG02-97ER54441.
\end{acknowledgments}


\end{document}